\documentstyle[aps,twocolumn,psfig]{revtex}

\renewcommand {\a}  {\alpha}

\newcommand   {\be} {\begin{equation}}
\newcommand   {\bea}{\begin{eqnarray}}

\newcommand   {\bgl}{\biggl}

\newcommand   {\bgr}{\biggr}

\newcommand   {\ct} {\cdot}
\newcommand   {\cts}{\cdots}
\renewcommand {\d}  {\partial}

\newcommand   {\dt} {\delta}
\newcommand   {\Dt} {\Delta}
\newcommand   {\ee} {\end{equation}}
\newcommand   {\eea}{\end{eqnarray}}
\newcommand   {\ep} {\epsilon}	

\newcommand   {\g}  {\gamma}
\newcommand   {\G}  {\Gamma}
\newcommand   {\h}  {\hat}
\renewcommand {\inf}{\infty}

\renewcommand {\l}  {\lambda}

\newcommand   {\lbl}{\label}
\newcommand   {\lts}{\ldots}

\newcommand   {\nn} {\nonumber}

\newcommand   {\ov} {\over}
\newcommand   {\p}  {\phi}

\newcommand   {\qq} {\qquad}
\newcommand   {\r}  {\rho}
\newcommand   {\ra} {\rightarrow}

\newcommand   {\sr} {\sqrt}

\newcommand   {\x}  {\xi}

\begin{document}
\draft
\title{Field Theoretical Analysis of On-line Learning of Probability Distributions\footnotemark}
\author{Toshiaki Aida}
\address{Department of Physics, Tokyo Institute of Technology,Oh-okayama, Meguro-ku, Tokyo, Japan}
\address{Tokyo Metropolitan College of Aeronautical Engineering, Minami-senjyu, Arakawa-ku, Tokyo, Japan}
\date{\today}
\maketitle
\begin{abstract}
On-line learning of probability distributions is analyzed from the field 
theoretical point of view. We can obtain an optimal on-line learning 
algorithm, since renormalization group enables us to control the number of 
degrees of freedom of a system according to the number of examples. 
We do not learn parameters of a model, but probability distributions 
themselves. Therefore, the algorithm requires no a priori knowledge 
of a model. 
\end{abstract}
\renewcommand{\thefootnote}{\fnsymbol{footnote}}
\footnotetext[1]{Copyright 1999 by The American Physical Society}


The methods of statistical mechanics have provided the framework to 
analyze learning and generalization by neural networks, and have achieved 
considerable success. This implies the fundamental connection between learning 
and statistical mechanical theories. 

For example, Bialek et al.\cite{BCS} made clear the batch learning problem of 
probability distributions from field theoretical point of view. 
They optimally controlled the number of degrees of freedom of a learning 
system through the scaling, which was naturally introduced by quantum field 
theory. 
It is crucial to control the number of degrees of freedom in learning 
problems. 
Learning a lot of features requires us to introduce a large number 
of degrees of freedom, while too many degrees of freedom increase 
ambiguities when we are given a small number of examples. This problem 
has been considered in various ways \cite{MYA}. 

Can we expect that the scaling leads to an optimal algorithm 
also in on-line learning problems? 
On-line learning has the advantages of less storage of data and less peak 
computation than batch learning. 
Some elaborated algorithms achieve the same 
optimal learning rate as the corresponding batch ones \cite{opt}. 
In this letter, we will show that the scaling analysis derives 
an optimal on-line learning algorithm, when we identify the number of examples 
with a scale parameter.

We observe a sequence of $D$-dimensional sample points 
${\bf x}_1 , {\bf x}_2 , \lts , {\bf x}_N$, which are drawn independently 
from an unknown probability distribution $Q^* ({\bf x})$. 
How can we infer the distribution, which is specified by an infinite number of 
parameters in general? 
A finite number of sample points do not enable us to 
determine $Q^* ({\bf x})$ uniquely, but only allows us 
to make a probabilistic description. 
We start from the probability of a distribution $Q({\bf x})$, which 
is given via Bayes' rule \cite{BCS}. 
\bea
&&P[ Q | {\bf x}_1 , \lts , {\bf x}_N ] 
= { P[ {\bf x}_1 , \lts , {\bf x}_N | Q ] \ P[ Q ] \ov 
P(  {\bf x}_1 , \lts , {\bf x}_N ) } , \nn \\ 
&& \qq \qq = { Q({\bf x}_1 ) \cts Q({\bf x}_N ) \ P[ Q ] \ov 
\int {\cal D}Q \ Q({\bf x}_1 ) \cts Q({\bf x}_N ) \ P[ Q ] } {. \! \! \!}
\lbl{eqn:Bayes}
\eea
Here, $P[Q]$ denotes a `prior distribution', 
which is 
a probability density in the space of probability distributions, 
and limits the function $Q$ to possible classes. 

Let us consider the properties of the prior distribution. 
Introducing a field valuable $\p ({\bf x})$ which takes a value 
$- \inf < \p < \inf$, we rewrite the distribution $Q({\bf x})$ 
as $Q({\bf x}) = e^{- \p ({\bf x}) } / l^D$. 
We note that $l$ is not a parameter but a unit length scale, which is 
different from the case of Ref.\cite{BCS}. 
It only plays a role to ensure the consistency of dimensions, and is 
fixed to be one in numerical studies. 
To make our learning problem well-posed, it is known that the prior 
distribution must not give ultraviolet divergences. 
Therefore, we choose such a prior 
distribution as imposes that the function $\p ({\bf x})$ has to be smooth 
enough ($2 \a > D$): 
\bea
&&P[ Q ]={1 \ov Z_0} \exp\bgl[- {l^{2 \a - D} \ov 2} \int d^D x \ 
(\d_x^\a \p )^2 \nn \\ 
&& \ -{1 \ov l^D} \int d^D x F(\p ) \bgr] 
\dt \bgl[ {1 \ov l^D} \int d^D x \ e^{- \p} - 1 \bgr] {. \! \! \!}
\lbl{eqn:prior}
\eea
Hereafter, we will use the
notation $\d_x^\a \equiv \d_{x_{i_1}} \cts \d_{x_{i_\a}}$, 
and mean by the repeated indices $i_1 , \lts ,i_\a $ the sums from $1$ to the 
dimension $D$. 
We note that we have introduced an unknown function $F( \p ({\bf x}))$, which 
will be determined later so that we may treat fluctuation effects in a 
renormalizable way. 
The delta function gives the constraint to normalize the probability 
distribution, and the factor $Z_0$ is a normalization constant. 

Putting the prior (\ref{eqn:prior}) into Bayes' rule (\ref{eqn:Bayes}), we 
obtain the partition function 
as $Z={1 \ov Z_0} \int {d \l \ov 2 \pi} \int {\cal D} \p \exp[- {1 \ov g} S(\p , \l )] $, where 
\bea
&&S={l^{2 \a - D} \ov 2} \int d^D x 
(\d_x^\a \p )^2 + {1 \ov l^D} \int d^D x F(\p ) \nn \\ 
&&+i\l \bgl[{1 \ov l^D}\int d^D x e^{- \p } -1 \bgr] 
+N\int d^D x P_N \p {. \! \! \!} 
\lbl{eqn:act}
\eea
The constant $g$ counts fluctuation effects, and is set to be one in 
numerical analysis. 
$P_N ({\bf x})$ is the distribution of sample data, defined 
as $P_N ({\bf x}) = {1 \ov N} \sum_{i=1}^N \dt^D ({\bf x} - {\bf x}_i )$. 
The action (\ref{eqn:act}) explicitly shows that we have no scale parameter 
other than the number of example data $N$. Therefore, we can conclude that 
the number $N$ sets the length scale to observe an unknown distribution and 
determines the behavior of a fluctuating system.

In order to proceed further, we will consider the asymptotic situation 
$N \ra \inf$. We expand the action (\ref{eqn:act}) around classical 
solutions, and perform the functional integration of $\p({\bf x})$. 
The classical solutions ${\h \p} ({\rm x})$ and $\h \l$ are defined 
by the following two equations. 
\bea
&&(-1)^\a l^{2 \a } \Dt^\a \h \p + F'
-i \h \l e^{- \h \p } 
= -N l^D P_N {, \! \! \!} 
\lbl{eqn:cl1} \\ 
&&{1 \ov l^D} \int d^D x \ e^{- \h \p ({\bf x})} = 1 \ . 
\lbl{eqn:cl2} 
\eea
If we apply the normalization condition (\ref{eqn:cl2}) 
to the integration of the equation (\ref{eqn:cl1}), 
we can find that $i \h \l =N+\int d^D x F'(\h \p ({\bf x}) )/ l^D$, 
which determines the mass term in the action (\ref{eqn:act}). 

It is straightforward to evaluate the functional integration of $\p({\bf x})$. 
We will calculate the leading corrections to the classical solutions. 
Then, it is enough to restrict ourselves to the quadratic order estimation. 
Furthermore, we will consider only the leading order effect 
in the $N \ra \inf$ limit. The integration up to quadratic 
terms is given by the ratio of functional determinants. We can 
evaluate it in a standard way \cite{C}. 
As a result, the effective action is found to be 
\be
S(\h \p , \h \l ) 
+gR \int d^D x \bgl({N \h Q \ov l^{2 \a - D} }\bgr)^{D \ov 2 \a} . 
\lbl{eqn:par2}
\ee
Here, the constant $R$ is defined as $R = 1 / (4 \pi )^{D/2 - 1} 4D \ \G ({D \ov 2}) \sin{{D \ov 2 \a} \pi} $.

Now that we have obtained a local correction term, we can determine the 
unknown function $F( \h \p )$. As is easily seen, a counter term can be 
produced only through the following definition of the function $F( \h \p )$: 
\bea
F( \h \p ) &=& k_0 \ e^{- {D \ov 2 \a} \h \p } \ , 
\lbl{eqn:defF} \\ 
k_0 &=& k \ - \ g R \ N^{D \ov 2 \a } \ . 
\lbl{eqn:renk}
\eea
Here, we have introduced a parameter $k$, which has the observed value 
at a scale $N$. 
Since a bare parameter $k_0$ 
does not depend on the scale $N$, 
we obtain the renormalization group equation which describes 
the scaling behavior of the parameter $k$. 
\be
{d k \ov d N} \ = \ g \ {D R \ov 2 \a } {1 \ov N^{ 1 - D/2 \a } } \ . 
\lbl{eqn:rgek}
\ee
As a result, we can find the scaling form of the parameter 
to be $k \sim g R N^{D/2 \a}$, which does not depend on the initial value 
of $k$. 

The scaling behavior is understood from the following point of view. 
In learning problems, the number of example data $N$ is considered to 
play a role of setting our observation scale. 
When we have received a small number of data points, we observe a signal 
source at a low resolution. The more examples we have received, at the higher 
resolution we can investigate the source. The signal source with a true 
probability distribution is a bare quantity which exists independently of 
our observation at a length scale. 
The renormalization group equation (\ref{eqn:rgek}) 
means that we should scale the parameter $k$ in order to respect the 
invariance of the true probability distribution under the change of the 
scale $N$. Therefore, such a scaling of the parameter $k$ is expected 
to lead to the optimal performance. 

On the other hand, we have the relation $k_0 = k$ if we do not consider 
the fluctuation effects in Bayesian setting. This is just the case of 
maximum likelihood estimation. Again, the invariance of $k_0$ under the 
change of $N$ requires that we should not vary the parameter $k$ 
with the number of examples $N$. It will be seen later that the scaling of the 
parameter $k$ plays a crucial role to achieve the optimal learning rate.

An on-line learning algorithm is how to change our hypothesis about an unknown 
distribution after receiving a new example. It is given by the recursion 
relation of the expectation value $<\p ({\bf x})>$ of the 
field $\p ({\bf x})$. In our approximation level, the expectation 
value is just the classical solution $\h \p ({\bf x})$. 
Therefore, an on-line learning algorithm is directly obtained from 
the variation of the equation (\ref{eqn:cl1}). 
\bea
&&\Dt<\p_N ({\rm x})> \sim {1 \ov g} \int d^D x' \ G_N ( {\bf x} , {\bf x}' ) 
\nn \\
&&\times \bgl[- \dt^D ({\bf x}' - {\bf x}_{N+1}) \ 
+ \ { \Dt (i \h \l )_N \ov l^D } e^{- <\p_N ({\bf x}')>} \nn \\ 
&& \ \ \ \ + {D \ov 2 \a l^D} {d k_N \ov d N} \ 
e^{- {D \ov 2 \a } <\p_N ({\bf x}')>} \bgr] . 
\lbl{eqn:dp} 
\eea
Hereafter, we will explicitly attach the number of examples $N$ to denote the 
scale. 
The learning rate $G_N ( {\bf x} , {\bf x}' )$ is a Green's function, 
and defines a local bin size $\x_N ({\bf x})$. 
\bea 
&&G_N ( {\bf x} , {\bf x}' ) \sim {g \ov l^{2 \a - D} } \ 
{ (-1)^{\a - 1} \ov (2 \pi )^{D \ov 2} \a \ r^{{D \ov 2} - 1} } \nn \\
&&\ \ \times \sum_{n=0}^{\a - 1} ( \g_n \xi_N^{-1} )^{ {D \ov 2} - 2 \a + 1 } 
K_{ {D \ov 2} - 1 } ( \g_n \xi_N^{-1} r ) {, \! \! \!} \\ 
\lbl{eqn:G} 
&&\x_N ({\bf x}) = l 
\bgl[ (i \h \l )_N e^{- <\p_N ({\bf x})>} \nn \\
&&\qq \ \ + \bgl( {D \ov 2 \a } \bgr)^2 k_N \ 
e^{- {D \ov 2 \a } <\p_N ({\bf x})>}
\bgr]^{- {1 \ov 2 \a } } {. \! \! \!}
\lbl{eqn:xi}
\eea 
Here, $r$ is the distance between the points {\bf x} and {\bf x}', and 
the function $K_{ {D \ov 2} - 1 }$ is a modified Bessel function of the 
second kind. $\g_n$ is defined to 
be $\g_n \equiv e^{ i (2 n + 1) {\pi \ov 2 \a } } e^{-i {\pi \ov 2}}$. 
The Green's function is real and regular for any $r \ge 0$. 

The first term on the right-hand side of (\ref{eqn:dp}) gives the effect of 
a new datum ${\bf x}_{N+1}$, which is necessarily local. The minus sign of the 
term ensures the increase of the distribution $<Q_N >\simeq e^{- <\p_N >}/l^D$ 
at the point. 
The second and third terms show the influence of the changes of the 
parameters $(i \h \l )_N$ and $k_N$ respectively. 
The Green's function smoothes them in the local length 
scale $\x_N ({\bf x})$. We can see from the equation (\ref{eqn:xi}) that the 
length scale should be small only where we find a lot of example data and 
when the number of examples $N$ is large. 
Thus, the bin size $\x_N ({\bf x})$ penalizes that we prepare too 
many degrees of freedom in order to reconstruct a probability distribution 
from given data. 

We note that we have not yet obtained the explicit form of 
the change $\Dt (i \h \l )_N$. We report having found 
that $\Dt (i \h \l )_N \sim 1 - D^2 k_N / 4 \a^2 N e^{ - (1 - {D \ov 2 \a }) <\p_N ({\bf x}_{N+1} )>}$, which is obtained from the variation 
of $(i \h \l )_N = N - D k_N / 2 \a \ct \int d^D x ( \h Q_N / l^{2 \a - D} )^{D / 2 \a}$.

In the rest, we will discuss the average performance of 
the algorithm. 
Since we consider the asymptotic situation, we may expand 
the algorithm around the true probability 
distribution $Q^* \equiv e^{- \p^* } / l^D$, which 
is an ultraviolet fixed point. Then, we define an error $\ep_N ({\bf x})$ 
as the difference $\ep_N ({\bf x}) \equiv <\p_N ({\bf x})> - \p^* ({\bf x})$, 
and will evaluate the dynamics of the error. 

We have expanded the Green's function in the convolution of the 
equation (\ref{eqn:dp}) around the fixed point, 
and found that it decays as $1/N$, which is known to be optimal in the 
on-line learning by neural networks \cite{O}.
Putting $<\p_N >=\p^* + \ep_N $ into the expansion, we can obtain 
the evolution equation of the error $\ep_N$. However, it requires 
care to be solved, since it has the $N$-dependence which is not extracted 
explicitly. The equation depends on a new datum ${\bf x}_{N+1}$ through the 
terms $\dt^D ({\bf x} - {\bf x}_{N+1})$ and $\Dt (i \h \l )_N$. 
The data dependence is averaged over many steps of evolutions, which is found 
from the relations \cite{BCS}: 
\bea
\sum_{i=1}^N \dt^D ({\bf x} - {\bf x}_{i+1}) 
&=& N Q^* ({\bf x}) + \sr{N} \r ({\bf x}) {, \! \! \!}
\lbl{eqn:del} \\ 
<< \r ({\bf x}) \r ({\bf x}') >> 
&=& Q^* ({\bf x}) \dt^D ( {\bf x} - {\bf x}' ) {. \! \! \!} 
\lbl{eqn:rho} 
\eea
Here, the function $\r ({\bf x})$ is a fluctuating density. 
The equation (\ref{eqn:del}) shows that we have to separate the data-dependent 
terms into a leading order and a sub-leading one. Therefore, we have to adopt 
the average $\int d^D x_{N+1} Q^* ( {\bf x}_{N+1} ) \Dt (i \h \l )_N$ as the 
leading order of $\Dt (i \h \l )_N$, and $\Dt (i \h \l )_N - \int d^D x_{N+1} 
\hfill\break 
Q^* ( {\bf x}_{N+1} ) \Dt (i \h \l )_N$ as the sub-leading one. 
Then, the leading order term $\ep_N^0 ({\bf x})$ of the large $N$ expansion of 
the error $\ep_N ({\bf x}) = \ep_N^0 ({\bf x}) + \ep_N^1 ({\bf x}) + \cts$ 
obeys the equation: 
\bea
& & N \ \ep_{N+1}^0 ({\bf x}) \ - \ (N-1) \ \ep_N^0 ({\bf x}) \nn \\ 
&=& 1-(Q^* )^{-1} \dt^D ({\bf x}-{\bf x}_{N+1} ) 
+g {D^2 R \ov 4 \a^2 N^{ 1 - {D \ov 2 \a} } } \nn \\ 
&\times&\bgl[ {1 \ov (l^D Q^* )^{ 1 - {D \ov 2 \a} } } 
-\int d^D x {Q^* \ov (l^D Q^* )^{ 1 - {D \ov 2 \a} } } \bgr] {. \! \! \!}
\lbl{eqn:eq0}
\eea
Summed up with $N$ changed $1$ to $N$, the equation (\ref{eqn:eq0}) gives the 
following result.
\bea
&&\ep_N^0 ({\bf x}) = - {1 \ov \sr{N} } \ Q^* ({\bf x})^{-1} \r ({\bf x}) 
+ g {DR \ov 2 \a N^{ 1 - {D \ov 2 \a} } } \nn \\ 
&&\times \bgl[ \ {1 \ov (l^D Q^* )^{ 1 - {D \ov 2 \a} } }
- \int d^D x {Q^* \ov (l^D Q^* )^{ 1 - {D \ov 2 \a} } } \bgr] {. \! \! \!}
\lbl{eqn:res0}
\eea
The sub-leading term $\ep_N^1 ({\bf x})$ is found to 
be $O(\log N /N) \ [\a = D]$ 
or $O(N^{ {D \ov 2 \a} - {3 \ov 2} }) \ [\a \neq D]$. 
However, we can prove that it does vanish 
if and only if $\a = D$. Then, the equation (\ref{eqn:res0}) is exact 
up to sub-leading order, and leads to the optimal error 
decay (\ref{eqn:sqerr}). This means that we can infer 
distributions most effectively, when we choose their candidates 
among $C^{2D}$-class functions in $D$-dimensions. 

From the equation (\ref{eqn:res0}), we can show that the average of the 
squared error $\ep_N ({\bf x})^2$ gives an universal result. 
This is consistent with the definition of the reparameterization invariant 
distance between two 
distributions $p({\bf x},{\bf \x} )$ and $p({\bf x},{\bf \x} + d {\bf \x} )$, 
which are specified by 
parameters $\x^i$ \cite{A}: $ds^2 = g_{ij} d \x^i d\x^j = E[{\d \ov \d \x^i} \log p \ {\d \ov \d \x^j} \log p ] d \x^i d \x^j = E[(d \log p)^2 ]$. 

We have found the quadratic error as 
\be
<< \int d^D x \ Q^* \ \ep_N{}^2 >> \ = \ {V_{\bf p} V_{\bf x} \ov (2 \pi)^D} 
{1 \ov N}. 
\lbl{eqn:sqerr}
\ee
Here, the coefficients $V_{\bf p}$ and $V_{\bf x}$ denote the volumes of 
momentum space and of coordinate one, respectively. 
The constant $V_{\bf p} V_{\bf x}$ gives the number of all the possible 
configurations of the system. 
Since it is independent of how to describe the system, we have reproduced 
the universal asymptotic behavior, which is well known in neural 
network models \cite{uni}. 
Thus, we have obtained the $1/N$ decay of the quadratic error up to 
sub-leading order, which is considered to be optimal 
as in the case of neural networks \cite{O}. 
We note that the optimal 
behavior (\ref{eqn:sqerr}) can be obtained, if and only if we develop the 
parameter $k_N$ according to the renormalization group 
equation (\ref{eqn:rgek}). 

\begin{figure}
 \psfig{file=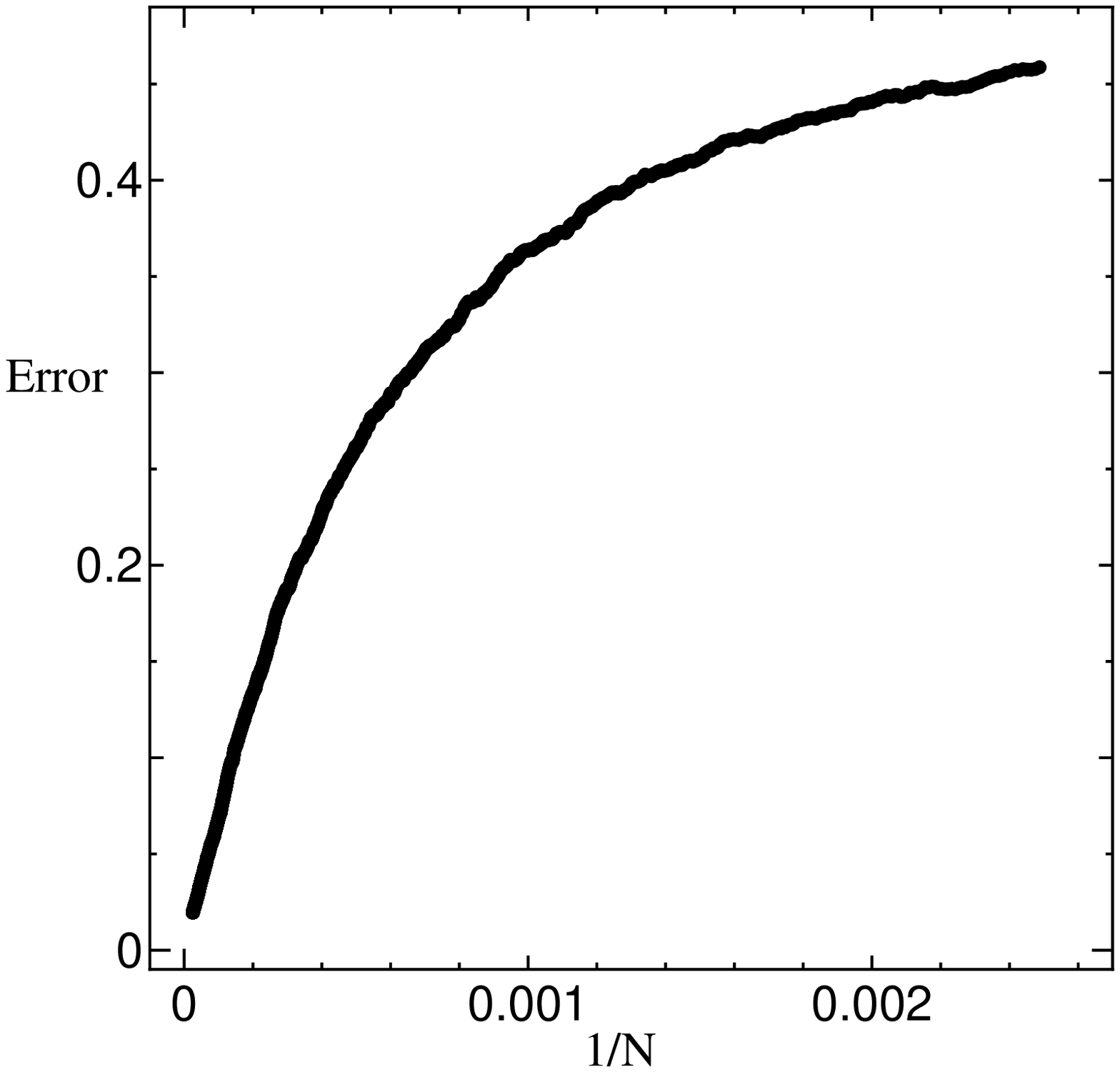,scale=0.35}
 \caption{The quadratic error (17) in learning the two-dimensional Gaussian 
distribution $Q^* ({\rm x}) = e^{- (x^2 + y^2 ) / 2} / 2 \pi \ [-1/2 \le x,y \le 1/2]$. 
We have performed the algorithm (10) with the area divided into $40\times40$ 
pieces.}
 \label{fig:gau}
\end{figure}

Figure \ref{fig:gau} shows the result of the numerical simulation of the 
algorithm (\ref{eqn:dp}), which is applied to the learning of 
a two-dimensional Gaussian distribution. 
It is obvious that the quadratic error is asymptotically linear to the inverse 
of the number of examples $1/N$. This ensures the analytical 
result (\ref{eqn:sqerr}). 

Finally, we will point out the relation to previous works and make 
concluding remarks. 
Bialek et al.\cite{BCS} gave an excellent field theoretical formulation of the 
batch learning of probability distributions. 
They controlled the number of degrees of freedom through the scaling of an 
infrared cut-off parameter $l$. 
Especially, they obtained the optimal scaling of a bin 
size $\x_N \propto N^{-1/(2 \a + D)}$, which includes the well-known 
result $\x_N \propto N^{-1/3}$ in one-dimensional 
space $D = \a = 1$ \cite{BC}. 
On the other hand, we have analyzed the on-line learning of probability 
distributions. 
In our case, renormalization group enables us to control a bin 
size (\ref{eqn:xi}) and leads to the optimal change of the 
hypothesis (\ref{eqn:dp}) through the scaling of the parameter $k_N$. We can 
find, from $(i \h \l)_N \sim N$ and $\a = D$, the optimal scaling of the bin 
size $\x_N \propto N^{-1/2D}$ in on-line learning scheme. 

Thus, renormalization group gives an on-line learning algorithm in a natural 
way, which we have proved to be optimal. Our discussion is so general that 
we may adopt it as a principle to derive optimal on-line 
learning algorithms. 
Furthermore, we could expect that the optimal error decay can be also 
achieved up to a desired order, if we extend the field theoretical 
analysis to the order.

The author would like to thank to H. Nishimori and Y. Kitazawa 
for their valuable comments and discussions. 
He also thanks J. Inoue for several useful comments. 
This work was partially supported by the Grant-in-Aid for Scientific Research 
No.10750056 from the Ministry of Education, Science and Culture, Japan. 

%
%
%
%
\end{document}